\documentclass[useAMS,usenatbib,onecolumn]{mn2e}
\usepackage{graphicx}
\usepackage{url}
\usepackage{color}

\newcommand{\cm}{cm$^{-1}$}

\newcommand{\state}{$X\,{}^2\Sigma^+$}
\newcommand{\ai}{\textit{ab initio}}

\newcommand{\eqref}[1]{(\ref{#1})}

\title{ExoMol line lists I: The rovibrational spectrum of BeH, MgH and CaH in the $X\,{}^2\Sigma^+$ state}

\date{\today}
\author[Yadin et al]{\large Benjamin Yadin, Thomas Veness, Pierandrea Conti,
Christian Hill, Sergei N. Yurchenko and Jonathan Tennyson \\
Department of Physics and Astronomy, University College London, Gower Street, WC1E 6BT London, UK}
\date{Accepted XXXX. Received XXXX; in original form XXXX}

\pagerange{\pageref{firstpage}--\pageref{lastpage}} \pubyear{2012}

\begin{document}

\label{firstpage}

\maketitle

\begin{abstract}

Accurate line lists for three molecules, BeH, MgH and CaH, in their ground
electronic states  are presented. These line lists are suitable for
temperatures relevant to exoplanetary atmospheres and cool stars (up to
2000K). A combination of empirical and \textit{ab initio} methods is used.
The rovibrational energy levels of BeH, MgH and CaH are computed using the
programs Level and DPotFit  in conjunction with `spectroscopic' potential
energy curves (PECs). The PEC of BeH is taken from the literature, while the
PECs of CaH and MgH are generated by fitting to the experimental transition
energy levels. Both spin-rotation interactions (except for BeH, for which
it is negligible) and non-adiabatic corrections  are explicitly taken into
account. Accurate line intensities are generated using newly computed
\textit{ab initio} dipole moment curves for each molecule using high levels
of theory. Full line lists of rotation-vibration transitions
for $^9$BeH, $^{24}$MgH, $^{25}$MgH, $^{26}$MgH and $^{40}$CaH
are made available in an electronic form as supplementary data to this article and at \url{www.exomol.com}.

\end{abstract}

\maketitle

\section{Introduction}

The ExoMol project aims to provide line lists of spectroscopic transitions
for key molecular species which are likely to be important in the atmospheres
of extrasolar planets and cool stars.  The aims, scope and methodology of the
project are summarised in the previous article \citep{ExoMol0}.
In the present paper, the first in a planned series, we present results for the
electronic ground state rotation-vibration spectra of three metal hydrides
BeH, MgH and CaH, denoted generically as XH. The line lists are explicitly
designed to extend to high temperatures. All three of these hydrides are of
astronomical interest.

The alkaline earth metal species BeH, MgH, and CaH share many similar
properties. Early spectroscopic data on these three species is captured by
\citet{HerzHub}. A summary of more recent studies, both spectroscopic and
astrophysical,  is presented for each molecule below.

\subsection{BeH}
BeH is one of simplest heteronuclear diatomic molecules, and hence it is a
strong contender for being observed in contexts such as exoplanetary
atmospheres, cool stars and in the interstellar medium. However there are only
very few astrophysical records of BeH; for example, a detection of
$A\,{}^2\Sigma^+$ $\to $ $X\,{}^2\Sigma^+$ emission lines of BeH
 in the sunspot umbra spectra by \cite{71Woxxxx.BeH} and \cite{08ShBaRa.BeH}.

BeH has been studied extensively by quantum chemists using different \ai\
methods. It is the simplest open-shell molecule, which made it very popular
as a benchmark object for testing new \ai\ approaches. {\it Ab initio}
studies on BeH are mostly limited by the basic spectroscopic and structural
constants. \cite{08PiSaVe.BeH}  reported very comprehensive
electronic-structure calculations on BeH with the potential energy and dipole
moments curves computed for a number of electronic states. A very accurate
\ai\ ground state PEC was  recently generated by \cite{11Koxxxx.BeH}.
\citet{06LeApCo.BeH} generated a `spectroscopic' potential energy curve (PEC)
of BeH obtained by fitting to the experimental values of
\cite{83CoDrSt.BeH,98FoFiBe.BeH,03ShTeBe.BeH}. This is currently the most
accurate PEC of BeH, and we employ it here in our production of the line list
for BeH.


Experimental work on BeH is not so extensive: most of experimental studies of
BeH have focused on electronic transitions
\citep{74DeCoxx.BeH,83CoDrSt.BeH,91ClCoxx.BeH,98FoFiBe.BeH}. Ground
electronic state rotation-vibration spectra were reported by
\citet{03ShTeBe.BeH}.

\subsection{MgH}
MgH is thought to be a likely candidate for interstellar observation but has
yet to be detected (\cite{98SaWhKaOh.MgH}). However its presence in stellar
spectra is well documented through observation of the
$A\,{}^2\Pi\!\to\!X\,{}^2\Sigma^+$ \citep{71Sotiro.MgH} and
$B^{\prime}\,{}^2\Sigma^+\!\to\!X\,{}^2\Sigma^+$ \citep{99WaHiLi.MgH}
transitions. These transitions have been used as an indicator for the
magnesium isotope abundances in the atmospheres of different stars
\citep{71LaMaPe.MgH,80ToLaxx.MgH,86LaMcxx.MgH,99WaHiLi.MgH,00GaLaxx.MgH,03DaYDaL.MgH}
from giants to dwarfs including the Sun. The same lines of MgH were used to
measure the temperature of stars (for example, by \cite{61Wyxxxx.MgH}),
surface gravity (for example, by \cite{85BeEdGu.MgH,93BoBexx.MgH}), stars'
metal abundance (for example, by \cite{78Coxxxx.MgH,80ToLaxx.MgH}),
gravitational acceleration (for example, by \cite{92BeSaxx.MgH}), to measure
the temperature and pressure \citep{90KuDeKn.MgH}, as well as for a deuterium
test \citep{08PaHaTe.MgH}. MgH is an important part of stellar
atmospheric models, such as, for example, PHOENIX \citep{01AlHaAl.MgH} which
incorporates the MgH line list of \cite{Kurucz}.

MgH has been studied in laboratories since 1929 (see \citep{29Pearse.MgH}), with
measurements of rovibrational transitions in the \state state given by
\citet{90ZiJeEv.MgH,93ZiBaAn.MgH} and most extensively by
\citet{04ShApGo.MgH}. Many experimentally measured spectral lines exist for
the $A\, {}^2\Pi \to X\, {}^2\Sigma^+$ and $B^\prime\, {}^2\Sigma^+ \to X\,
{}^2\Sigma^+$ transitions \citep{85BeBlBr.MgH,07ShHeLe.MgH,11ShBexx.MgH},
although the ground state transitions have only been observed in the infrared
up to $v=4$. Experimental ground $X{}^2\Sigma^+$ state spectra were reported
 by \cite{86LeZiEv.MgH,88LeDeDe.MgH,90ZiJeEv.MgH,93ZiBaAn.MgH},
which also include high-resolution millimeter-wave studies of the pure
rotational transitions with the hyperfine structure resolved.

On the \ai\ side, the calculations of the potential energy curves  and dipole
moment curves (DMCs) of MgH by \citet{78SaKiLi.MgH} made a very important
contribution to the development of the theory for this molecule. The results
of this study have been used in many spectroscopic applications including the
$^{24}$MgH line list from the UGAMOP  group
\citep{03WeScSt.MgHline,03WeScSt.MgHcontinuum}. Very recently, accurate \ai\
calculations have been performed using significantly higher levels of theory
\citep{09MePuSo.MgH,10GuSpFe.MgH}, with PECs and DMCs of MgH reported.
Important theoretical work for the present study is by \cite{07ShHeLe.MgH},
where an accurate `spectroscopic' PEC of MgH was obtained by fitting to all
existing experimental transitions  for this molecule. We use this PEC as an
initial approximation in our fits.

The first complete MgH line list was computed by Kurucz (see \cite{Kurucz}).
This line list, however, overestimates the opacity of MgH when included into
the atmospheric models due to presence of nonexistent levels (see discussion
by \cite{03WeScSt.MgHline}). Kurucz's line list does not include the ground
rotation-vibration transitions and thus cannot be compared to the results of
the present work. The UGAMOP line lists
\citep{03WeScSt.MgHcontinuum,03WeScSt.MgHline,03WeStKi.MgH} mentioned above
contain 23~315 transition energies and oscillator strengths over the
wavelength range 0 -- 32,130 \cm, for all possible allowed transitions from
the ground electronic state vibrational levels with $v^{\prime\prime}\le 11$.

\subsection{CaH}
The main astrophysical interest in CaH is because of the spectra of T Tauri stars
\citep{04Dixxxx.CaH}. The transition lines of CaH were used in
identifications and studies of  different dwarfs
\citep{93BaScGr.CaH,97ReGiCo.CaH,03BuKiLe.CaH,04McKiMc.CaH,07ReHoHa.CaH},
galactic disks \citep{97ReGiCo.CaH}, molecular clouds \citep{98SaWhKa.CaH},
and sun spots \citep{06BeFlRa.CaH}. One of the most important spectroscopic
features of the CaH spectra of T and L dwarfs  is the 6750 -- 7050 nm band
covering the $A-X$ $(0,0$) rotational transitions
\citep{03BuKiLe.CaH,07ReHoHa.CaH}.

CaH has been studied in the laboratory since \citet{27Hulthe.CaH} and
\citet{35WatWeb.CaH}.  Much experimental work has been done on the electronic
transitions of CaH
\citep{03WeStKi.CaH,11RaTeGo.CaH,95LeJexx.CaH,06ChGeSt.CaH}, but only a few
studies  exist on the ground state rovibrational spectrum. The most important
experimental studies of CaH for the current work include the Fourier
transform emission spectra of the   $E\, {}^2\Pi-X\, {}^2\Sigma^+$ bands
\citep{11RaTeGo.CaH}  and of four ro-vibrational bands in the $X\,
{}^2\Sigma^+$ state of CaH and CaD \citep{04ShWaGo.CaH}. These works provided
very accurate experimental measurements of rovibronic frequencies with a very
thorough analysis. Other experimental work include spectra and analysis of
the $A-X$ and $B-X$ systems of CaH  and CaD
\citep{74BeKlxx.CaH,84Maxxxx.CaH,76BeKlMa.CaH,81KlMaxx.CaH,03GaAlSk.CaH,02PeSkUr.CaH};
ultraviolet-absorption spectra of the C, D, K, L, G, J, and M bands of CaH
\citep{74KaLiRa.CaH,74KaLixx.CaH,76KaLixx.CaH,79BeHeJo.CaH,81KaLixx.CaH};
magnetic hyperfine structure of the $A-X$ and $B-X$ spectra of CaH
\citep{87StMeAl.CaH,06ChGeSt.CaH}; and the IR \citep{89PeLeDe.CaH,93FrPixx.CaH}
and millimeter-wave \citep{93BaAnZi.CaH,93FrOhCo.CaH} transitions within the
ground electronic state.

From a very large number of \ai\ studies of  CaH we select here only most
recent and comprehensive works. The potential energy and dipole moment curves
of CaH were computed using different levels of theory
(\cite{92BoDaEl.CaH,95LeJexx.CaH,06HoUrxx.CaH,07KeMaxx.CaH}). In most cases a
number of equilibrium constants, including $r_{\rm e}$, $\omega_{\rm e}$ and
$D_{\rm e}$, as well as the permanent dipole moments $\mu_e$ were reported for
one or several low lying states. To our knowledge there are no experimental
absolute intensity measurements of CaH. However, a few lifetime measurements
for a number of low lying vibrational states $v=0,1,2$ of the $X$, $A$, and
$B$ states were used to estimate the corresponding oscillator strengths or
Einstein coefficients \citep{82KlMaNy.CaH,96BeEkKe.CaH,09LiPaSj.CaH}. The
permanent electric dipole moments of CaH were obtained experimentally for a
few low electronic states using the Stark-effect
\citep{04StChG1.CaH,08ChStTi.CaH}.

As far as the theoretical ro-vibronic studies on CaH are concerned, there
have been attempts to refine PECs by fitting to available experimental
energies or frequencies \citep{88Maxxxx.CaH,98Uexxxx.CaH,00UrGoGo.CaH}.
However, previous ro-vibronic models did not include a proper description
of different important couplings between electronic states. Therefore only
the ground electronic state could be  reasonably well described in this
approach, for example by \cite{98Uexxxx.CaH} who also included the
Born-Oppenheimer breakdown (BOB) effect. None of these studies   take account
of spin-rotational splittings. \citet{02CaElNu.CaH} presented an accurate
model with an explicit coupling of four lowest electronic states as well as
the BOB effect using previously published \ai\ PECs by \cite{88Maxxxx.CaH}.
However since these PECs are of \ai\ (not experimental) accuracy, the
resulting energies were only good enough to validate the models used, but not
to predict the corresponding frequencies of CaH and CaD.

The most comprehensive line list for $^{40}$CaH was reported by \cite{03WeStKi.CaH},
who produced a complete list of CaH rovibrational levels and oscillator
strength covering all  possible allowed and bound transitions (89~970) in the
frequency range up to 35000 \cm\ involving the $X\,{}^2\Sigma^+$, $A\,{}^2\Pi$,
$B/B^\prime\, {}^2\Sigma^+$, $C\, {}^2\Sigma^+$, $D\, {}^2\Sigma^+$, and  $E\, ^2\Pi$
states. This line list should be good up to at least 2000~K and is also
included into the UGAMOP data base. In the production of this data the
authors used the most recent set of theoretical PECs and DMCs, with
adjustments to account for experimental dissociation energies and asymptotic
limits. However the model used to compute the frequencies was not very
accurate: (i) no fine structure, such as spin-rotation, or non-adiabatic
effects, were taken into account; (ii) the potential energy curves were not
adjusted to the experiment thus the overall accuracy is no better than 10 to 30 \cm.

In this work we present detailed line lists for
BeH, $^{24}$MgH, $^{25}$MgH, $^{26}$MgH and $^{40}$CaH
considering only the rotation-vibration transitions in the \state\ state. At
temperatures up to about 2000~K, any population of excited electronic states
will be insignificant compared to the population of the ground state. Taking
typical values for energies in the ground and first excited states of XH
\citep{08PiSaVe.BeH,10GuSpFe.MgH,07KeMaxx.CaH}, the thermalised, relative
population of the lowest rovibrational level of the first excited electronic
state to that of the ground state for BeH, MgH and CaH can be estimated to be
$2\times10^{-6}$, $2\times 10^{-5}$ and $1\times 10^{-4}$ respectively at
2000~K. In practice the excited states are all close to or above the dissociation
limits for the species so dissociative processes will further deplete the
excited state populations.

\section{Method}

For the species XH we solve the Schr\"{o}dinger equation allowing for Born-Oppenheimer breakdown effects as given by \cite{07ShHeLe.MgH} using
two programs, Level 8.0 \citep{lr07} and  DPotFit 1.1 \citep{dpotfit}.

We did not generate new {\it ab initio} PECs  for BeH and MgH. Full sets
of potential parameters representing very accurate PECs for these molecules
were already available from \cite{06LeApCo.BeH} and \cite{07ShHeLe.MgH},
respectively. {\it Ab initio} potential energy curves for CaH were generated
using multiple levels of theory, beginning with a simple Hartree-Fock (HF)
method and progressing to complete active space self consistent field
(CASSCF)  and multi-reference configuration interaction (MRCI)
\citep{Knowles-MRCI-1988}. The best potential, with no unphysical kinks and a
shape in agreement with previous works \citep{07KeMaxx.CaH,04ShWaGo.CaH},
used the MRCI/aug-cc-pV5Z level of theory. All electronic structure
calculations were performed with the quantum chemistry package
Molpro~2010~\citep{MOLPRO}.

\begin{figure}
\begin{center}
\includegraphics{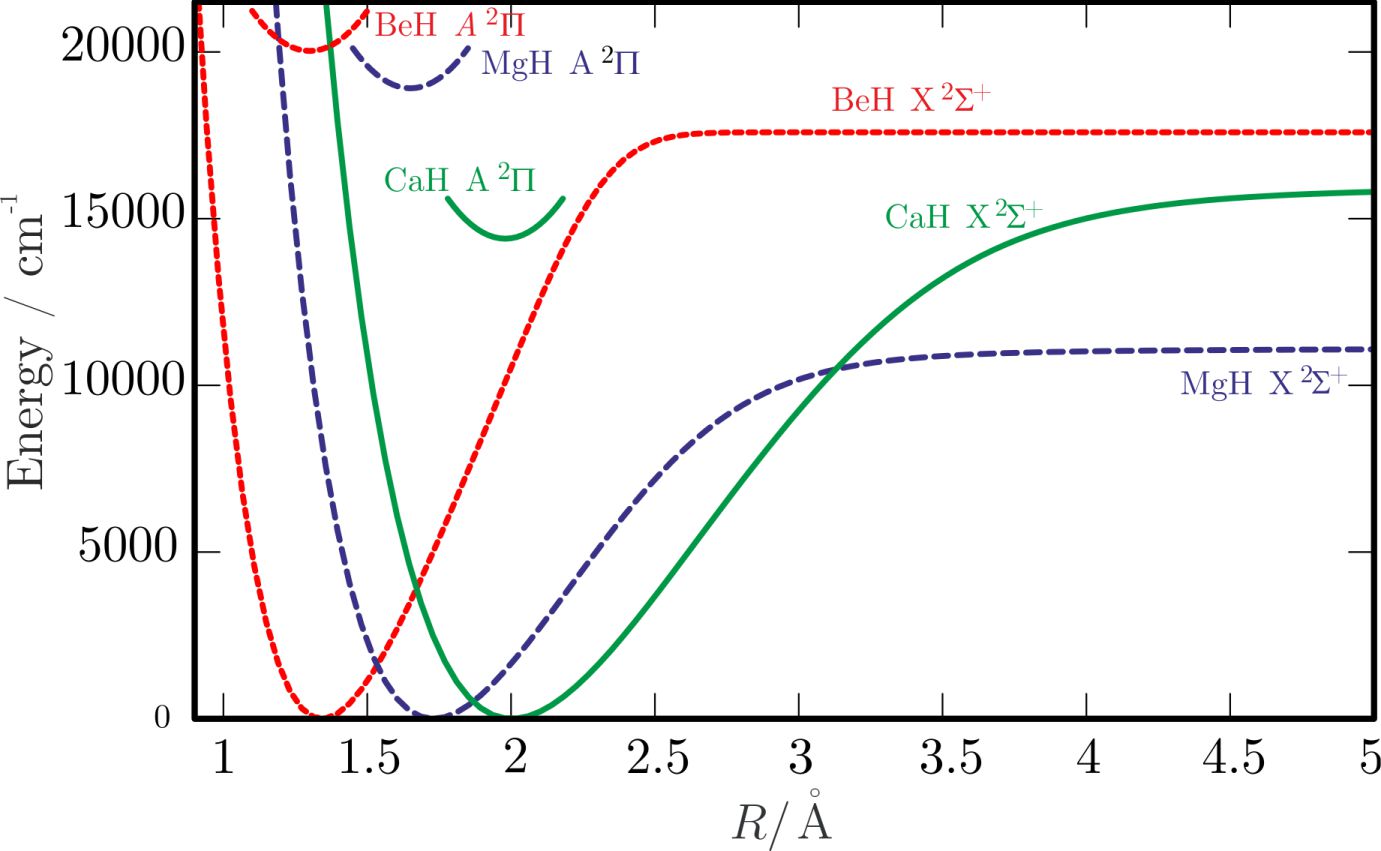}
\caption{Potential energy curves of CaH (solid line), MgH (dashed line), and BeH (dotted line) in their ground electronic states. The associated first excited
states are also indicated.}
\label{fig:allpots}
\end{center}
\end{figure}

\begin{figure}
\begin{center}
\includegraphics{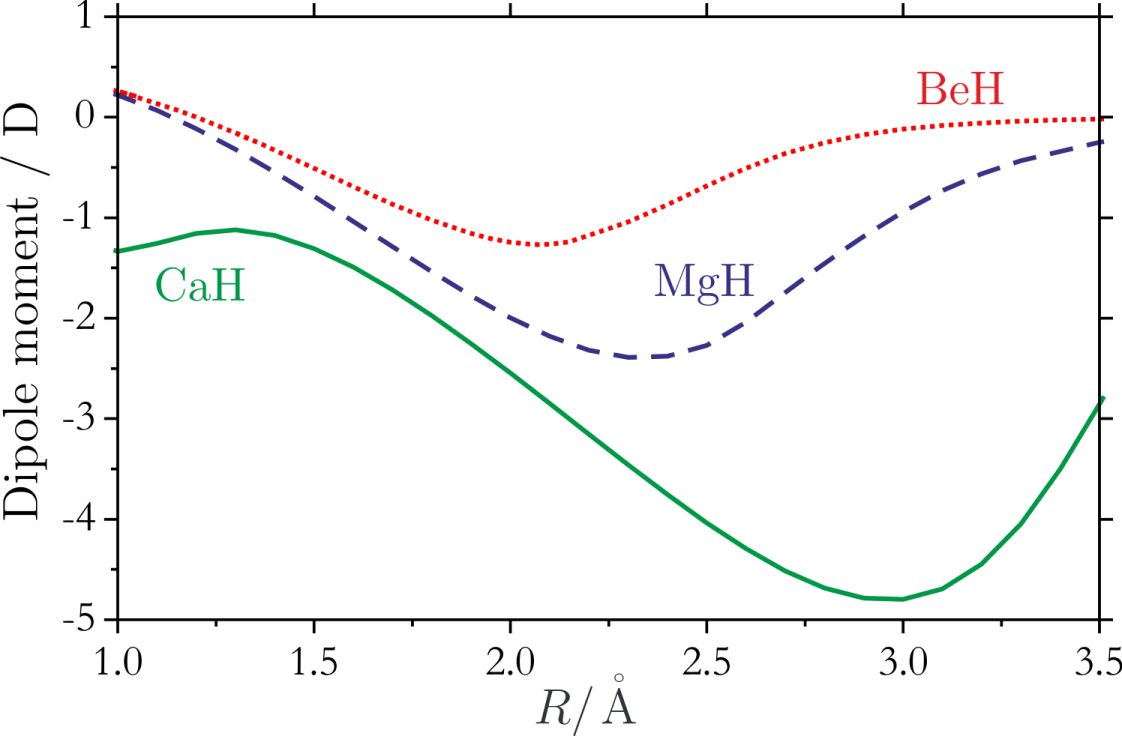}
\caption{The \ai\ dipole moment curves of BeH, MgH and CaH in their ground electronic states selected for the line list production.}
\label{fig:CaHdipole1}
\end{center}
\end{figure}

The DMCs for all three molecules were generated \ai\ employing Molpro and
different levels of theory.  The best DMCs for each molecule (shown on
Fig.~\ref{fig:CaHdipole1}) were selected based on their smoothness and
agreement with published curves
\citep{78SaKiLi.MgH,08PiSaVe.BeH,95LeJexx.CaH,10GuSpFe.MgH}. The DMCs were
calculated using a finite field rather than as an expectation value as this
is known to give more accurate results \citep{jt475}.

For BeH the final DMC was computed using the aug-cc-pwCV5Z basis set and the
MRCI method on a grid of 45 geometries distributed between 1.3 and 10~\AA.
Our equilibrium dipole moment is $-$0.224~D at $r_{\rm e} = 1.342$~\AA. This
can be compared to $-$0.199~D \citep{08PiSaVe.BeH} for the same value of
$r_{\rm e}$.

For MgH the final DMC was calculated using RCCSD(T)
\citep{Knowles-RCCSD-1993,Knowles-RCCSD-2000} in conjunction with the basis
sets cc-pCV5Z \citep{Woon-cc-pCVXZ-1995} for Mg and aug-cc-pV5Z
\citep{Woon-aug-cc-pVXZ-1993}  for H and all electrons correlated. The good
performance of the RHF-CCSD methods in the case of MgH was noted by
\cite{95LiPaxx.MgH}.  At $r_{\rm e} = 1.730$~\AA\ we obtain an equilibrium
value of the dipole moment $\mu_{\rm e}$ of $-$1.371~D. This can be compared
to the CCSD(T) value from \cite{98SeCaxx.MgH}, $-$1.356~D. Our DMC agrees
surprisingly well with dipole curves computed at the SDTCI level of theory by
\citet{78SaKiLi.MgH}.

The  DMC for CaH was also calculated using RCCSD(T)/cc-pCV5Z with all
electrons correlated. \citet{06HoUrxx.CaH} demonstrated importance of the
electron correlation contribution (0.44~D or 17.4\% of the total value),
while the relativistic and complete basis set corrections are relatively
small (about 0.009~D and 0.007~D, respectively).  Our equilibrium value of
the RCCSD(T)/cc-pCV5Z dipole moment is 2.55~D, which can be compared to the
experimentally derived value of the permanent dipole moment 2.53(3)~D
measured for the ground $X\,{}^2\Sigma^+(v=0)$ state with the optical Stark
effect \citep{08ChStTi.CaH}. \citet{06HoUrxx.CaH} recently obtained an \ai\ value of 2.463~D for $\mu_{\rm e}$
using a high level of theory.

In all our calculations the \ai\ DMC grid points were used directly in Level.

\subsection{Fitting the potential}

%

In the case of BeH we employed the spectroscopic PEC of
\citet{06LeApCo.BeH} consisting of an extended Morse oscillator (EMO) function corrected by BOB. We
could not precisely reproduce the ro-vibrational energies reported in this
paper using these potential parameters  and Level~8.0, see columns~II and III
of Table~\ref{tab:BeH:energies}. The deviations in the vibrational term
values are rather small, except for $v=11$ which is about 10~\cm, therefore
we decided to employ the PEC from \cite{06LeApCo.BeH} in the line list
calculations without any additional refinements.

In the case of the  PEC of MgH we experienced a similar problem. Using the
potential parameters from \cite{07ShHeLe.MgH} in conjunction with the Morse
Long Range potential (MLR) \citep{MLRpaper} in the DPotFit calculations we
could not reproduce frequencies obtained by \cite{07ShHeLe.MgH} using the
same program. Considering importance of MgH for astrophysical applications we
decided to refine \citet{07ShHeLe.MgH}'s parameters through fits to
the experimental data and using the same program (i.e. DPotFit
1.1) to guarantee the quality of the line list produced. In these fits, to
avoid dealing with the high electronic states, we constructed a set of `fake'
transitions within the ground electronic states from the experimental
transitions $A\, {}^2\Pi \to X\, {}^2\Sigma^+$ and $B\, {}^2\Sigma^+ \to X\,
{}^2\Sigma^+$ using the method of the combination differences. As a result
our `experimental' set included 1302 transition frequencies, 380 of which
were actual experimental, while the rest were `experimentally' derived
values. Some of the lines, especially from highly excited vibrational
excitations, had a greater uncertainty and therefore were given a lower
weighting. Despite their relatively low quality, these data are important to
fix the behaviour of the potential at large inter-molecular distances and
hence to accurately predict the higher-energy states.  The fit was found to
be unstable unless the dissociation energy $D_e$ was held fixed -- the best
available value of 11104.7 \cm\ from \citet{07ShHeLe.MgH} was chosen.

The fit of the MgH potential achieved a root-mean-square (rms)
error of 0.0012 \cm\ for the 380 experimental $X\,{}^2\Sigma^+ - X\,{}^2\Sigma^+$ lines with a maximum difference of 0.0077 \cm, showing that
these lines are reproduced to high accuracy despite the inclusion of extra
lines in the fit. Compared with the experimental line list (6076 lines)
calculated by combination differences, the rms error is 0.059 \cm\ and the
maximum difference is 0.35 \cm. The `experimentally' derived lines show an
inherent uncertainty of up to about 0.2 \cm.

To our knowledge there is no potential for CaH in the literature of a
satisfactory accuracy. Therefore we built a new `spectroscopic' PEC for this
molecule using the MLR potential. Fitting to the experimental data requires a
good initial estimate for the PEC, for which our MRCI-Q/aug-cc-pV5Z potential
was used. The corresponding MLR potential parameters were obtained through a
fitting to the \ai\ energies (see above) distributed over the geometry range
from 0.7 to 10.0 \AA.  The potential was then refined by fitting to a list of
426 transition frequencies of $^{40}$CaH  by \cite{04ShWaGo.CaH} and \citet{11RaTeGo.CaH}.
We obtained higher accuracy fits if $D_e$ was allowed to vary. Our value of
$D_e$ = 15~923 \cm\ ended up somewhat higher than the experimental value
14~800 \cm\ \citep{HerzHub}. The CaH fit achieved an rms of 0.001 \cm\ for
426 lines, the most complete data set we could build for fitting against.
It should also be noted that no differences exceeded 0.0087 \cm.

For CaH, the most important change from the initial \ai\ parameters is
attributed to the equilibrium geometry. The initial parameters of MgH, being
already empirical, however, were significantly less affected by the fitting
process.

To illustrate the quality of these PECs, Tables \ref{tab:BeH:energies} --
\ref{tab:CaH:energies} compare our vibrational energies, $T_v$, and rotational
constants, $B_v$, with the corresponding `experimental'  (when
available) and other theoretical values. In case of MgH and CaH we list the
UGAMOP theoretical values
\citep{03WeScSt.MgHcontinuum,03WeScSt.MgHline,03SkWeSt.MgH,03WeStKi.MgH,03WeStKi.CaH},
while for BeH \ai\ spectroscopic constants of \cite{06LeApCo.BeH} are given.
When comparing to our energies it should remembered that the UGAMOP energies do not
include the spin-rotational splitting. Our results exhibit a significant
improvement  relative to the UGAMOP values for MgH and CaH.

\begin{table}
\caption{A comparison of the theoretical and experimental spectroscopic constants of BeH in \cm. }
\label{tab:BeH:energies} \footnotesize
\begin{center}
\begin{tabular}{lrrr@{}lr}
\hline
            &  Obs.$^a$      &        Calc.$^b$  &  \multicolumn{2}{c}{Obs.$-$Calc.}              & \cite{06LeApCo.BeH}   \\
         \hline
$T_1$                     &      1986.4442 &            1986.4176 &   $      0.0266 $  &   $                $  &      1986.4163   \\
$T_2$                     &      3896.8785 &            3896.8714 &   $      0.0071 $  &   $                $  &      3896.8712   \\
$T_3$                     &      5729.2861 &            5729.2618 &   $      0.0243 $  &   $                $  &      5729.2602   \\
$T_4$                     &      7480.4528 &            7480.3400 &   $      0.1128 $  &   $                $  &      7480.3383   \\
$T_5$                     &      9145.2834 &            9145.2533 &   $      0.0301 $  &   $                $  &      9145.1320   \\
$T_6$                     &     10716.2804 &           10716.6292 &   $     -0.3488 $  &   $                $  &     10716.1629   \\
$T_7$                     &      12182.323 &           12183.0505 &   $     -0.7275 $  &   $                $  &     12182.2070   \\
$T_8$                     &      13525.889 &           13526.6563 &   $     -0.7673 $  &   $                $  &     13525.7882   \\
$T_9$                     &      14717.860 &           14718.8262 &   $     -0.9662 $  &   $                $  &     14718.0819   \\
$T_{10}$                  &      15709.040 &           15710.4672 &   $     -1.4272 $  &   $                $  &     15709.3839   \\
$T_{11}$                  &                &           16402.0939 &   $             $  &   $                $  &     16412.7734   \\
$B_0$                     &     10.1648880 &            10.165644 &   $       -7.56 $  &   $ \times10^{-4}  $  &      10.165715   \\
$B_1$                     &      9.8554335 &             9.855734 &   $       -3.01 $  &   $ \times10^{-4}  $  &       9.855827   \\
$B_2$                     &      9.5417271 &             9.541630 &   $        0.98 $  &   $ \times10^{-4}  $  &       9.541718   \\
$B_3$                     &       9.220768 &             9.220113 &   $        6.55 $  &   $ \times10^{-4}  $  &       9.220280   \\
$B_4$                     &       8.886639 &             8.886566 &   $        0.73 $  &   $ \times10^{-4}  $  &       8.886738   \\
$B_5$                     &       8.533209 &             8.534221 &   $      -10.12 $  &   $ \times10^{-4}  $  &       8.534429   \\
$B_6$                     &       8.152609 &             8.152146 &   $        4.63 $  &   $ \times10^{-4}  $  &       8.153772   \\
$B_7$                     &       7.729457 &             7.722324 &   $       71.33 $  &   $ \times10^{-4}  $  &       7.728265   \\
$B_8$                     &         7.2309 &             7.214164 &   $      166.86 $  &   $ \times10^{-4}  $  &       7.225306   \\
$B_9$                     &         6.5921 &             6.570639 &   $      214.91 $  &   $ \times10^{-4}  $  &       6.579663   \\
$B_{10}$                  &         5.6984 &             5.662832 &   $      355.38 $  &   $ \times10^{-4}  $  &       5.667413   \\
$B_{11}$                  &                &                      &   $             $  &   $                $  &       4.268157   \\
\hline
\end{tabular}

$^a$ Experimentally derived by \cite{98FoFiBe.BeH}; $^b$ Calculated, this work.

\end{center}

\end{table}

\begin{table}
\caption{A comparison of the theoretical and experimental spectroscopic constants of $^{24}$MgH in \cm. }
\label{tab:MgH:energies} \footnotesize
\begin{center}
\begin{tabular}{lrrr@{}lr}
\hline
            &     Obs.$^a$   &       Calc.$^b$  &  \multicolumn{2}{c}{Obs.$-$Calc.}              & UGAMOP   \\
         \hline
$T_1$                     &     1431.97786 &           1431.9777 &   $  0.0002 $  &   $                $  &      1423.0023   \\
$T_2$                     &     2800.67807 &           2800.6770 &   $  0.0011 $  &   $                $  &      2784.3944   \\
$T_3$                     &     4102.32975 &           4102.3281 &   $  0.0017 $  &   $                $  &      4081.3781   \\
$T_4$                     &      5331.3892 &           5331.3867 &   $  0.0025 $  &   $                $  &      5308.9675   \\
$T_5$                     &      6479.6562 &           6479.6547 &   $  0.0015       $  &   $                $  &      6459.3356   \\
$T_6$                     &      7534.8137 &           7534.8076 &   $  0.0061       $  &   $                $  &      7521.0069   \\
$T_7$                     &      8477.9997 &           8477.9959 &   $ 0.0038        $  &   $                $  &      8477.2820   \\
$T_8$                     &      9279.6527 &           9279.6436 &   $  0.0091       $  &   $                $  &      9299.8872   \\
$T_9$                     &      9892.7243 &           9892.7201 &   $  0.0042       $  &   $                $  &      9929.2662   \\
$T_{10}$                  &     10249.4074 &          10249.4050 &   $  0.0020       $  &   $                $  &     10150.2699   \\
$T_{11}$                  &       10352.25 &          10352.2503 &   $   0.00      $  &   $                $  &     10231.8603   \\
$B_0$                     &     5.73650768 &           5.7365078 &   $   -0.01 $  &   $ \times10^{-5}  $  &                  \\
$B_1$                     &     5.55528801 &          5.55528804 &   $  -0.003 $  &   $ \times10^{-5}  $  &                  \\
$B_2$                     &       5.367512 &          5.36756316 &   $   -5.12 $  &   $ \times10^{-5}  $  &                  \\
$B_3$                     &       5.169725 &          5.16981489 &   $   -8.99 $  &   $ \times10^{-5}  $  &                  \\
$B_4$                     &       4.956539 &          4.95667678 &   $  -13.78 $  &   $ \times10^{-5}  $  &                  \\
$B_5$                     &        4.71938 &          4.71971001 &   $  -33.00 $  &   $ \times10^{-5}  $  &                  \\
$B_6$                     &        4.44431 &          4.44473868 &   $  -42.87 $  &   $ \times10^{-5}  $  &                  \\
$B_7$                     &         4.1072 &          4.10695883 &   $   24.12 $  &   $ \times10^{-5}  $  &                  \\
$B_8$                     &        3.65877 &          3.65971997 &   $  -95.00 $  &   $ \times10^{-5}  $  &                  \\
$B_9$                     &        3.00781 &          3.00747254 &   $   33.75 $  &   $ \times10^{-5}  $  &                  \\
$B_{10}$                  &         1.9687 &           1.9684569 &   $   24.31 $  &   $ \times10^{-5}  $  &                  \\
$B_{11}$                  &          0.884 &          0.88689093 &   $ -289.09 $  &   $ \times10^{-5}  $  &                  \\
\hline
\end{tabular}

$^a$ Experimentally derived by \citep{07ShHeLe.MgH}; $^b$ Calculated, this work.

\end{center}

\end{table}

\begin{table}
\caption{A comparison of the theoretical and experimental spectroscopic constants of $^{40}$CaH in \cm. }
\label{tab:CaH:energies}

\footnotesize
\begin{center}
\begin{tabular}{lrrr@{}lr}
\hline
                          &  Obs.$^a$ &         Calc.$^b$     &     \multicolumn{2}{c}{Obs.$-$Calc.}     &    UGAMOP \\
\hline
$T_1$                     &     1260.12775 &            1260.1275 &   $  0.0003 $  &   $                $  &      1248.1020   \\
$T_2$                     &     2481.99888 &            2481.9986 &   $  0.0003 $  &   $                $  &      2459.1800   \\
$T_3$                     &      3665.4141 &            3665.4135 &   $  0.0006 $  &   $                $  &      3634.0060   \\
$T_4$                     &      4809.9464 &            4809.9453 &   $  0.0011 $  &   $                $  &      4772.1695   \\
$T_5$                     &                &            5914.8483 &   $         $  &   $                $  &      5873.3462   \\
$T_6$                     &                &            6979.2865 &   $         $  &   $                $  &      6936.7620   \\
$T_7$                     &                &            8002.3180 &   $         $  &   $                $  &      7957.0594   \\
$T_8$                     &                &            8982.6564 &   $         $  &   $                $  &      8926.4122   \\
$T_9$                     &                &            9918.3259 &   $         $  &   $                $  &      9836.9332   \\
$T_{10}$                  &                &           10806.3117 &   $         $  &   $                $  &     10680.8451   \\
$T_{11}$                  &                &           11642.2418 &   $         $  &   $                $  &     11447.2650   \\
$T_{12}$                  &                &           12420.0739 &   $         $  &   $                $  &     12118.8518   \\
$T_{13}$                  &                &           13131.7349 &   $         $  &   $                $  &     12676.5357   \\
$T_{14}$                  &                &           13766.6602 &   $         $  &   $                $  &     13121.7989   \\
$T_{15}$                  &                &           14311.2476 &   $         $  &   $                $  &     13461.1465   \\
$T_{16}$                  &                &           14748.4857 &   $         $  &   $                $  &     13696.1992   \\
$T_{17}$                  &                &           15059.0529 &   $         $  &   $                $  &                  \\
$B_0$                     &      4.2286902 &           4.22868360 &   $    0.66 $  &   $ \times10^{-5}  $  &                  \\
$B_1$                     &       4.131722 &           4.13171966 &   $    0.23 $  &   $ \times10^{-5}  $  &                  \\
$B_2$                     &      4.0342454 &           4.03424554 &   $   -0.01 $  &   $ \times10^{-5}  $  &                  \\
$B_3$                     &       3.935887 &           3.93589433 &   $   -0.73 $  &   $ \times10^{-5}  $  &                  \\
$B_4$                     &       3.836122 &           3.83614878 &   $   -2.68 $  &   $ \times10^{-5}  $  &                  \\
$B_5$                     &                &           3.73452399 &   $         $  &   $                $  &                  \\
$B_6$                     &                &           3.63062100 &   $         $  &   $                $  &                  \\
$B_7$                     &                &           3.52382050 &   $         $  &   $                $  &                  \\
$B_8$                     &                &           3.41293766 &   $         $  &   $                $  &                  \\
$B_9$                     &                &           3.29601845 &   $         $  &   $                $  &                  \\
$B_{10}$                  &                &           3.17024576 &   $         $  &   $                $  &                  \\
$B_{11}$                  &                &           3.03183969 &   $         $  &   $                $  &                  \\
$B_{12}$                  &                &           2.87582915 &   $         $  &   $                $  &                  \\
$B_{13}$                  &                &           2.69557438 &   $         $  &   $                $  &                  \\
$B_{14}$                  &                &           2.48189083 &   $         $  &   $                $  &                  \\
$B_{15}$                  &                &           2.22158167 &   $         $  &   $                $  &                  \\
$B_{16}$                  &                &           1.89536199 &   $         $  &   $                $  &                  \\
$B_{17}$                  &                &           1.47691639 &   $         $  &   $                $  &                  \\
\hline
\end{tabular}

$^a$ Experimentally derived by \cite{04ShWaGo.CaH}; $^b$ Calculated, this work.

\end{center}

\end{table}

\section{Line list calculations}

The procedure described above was used to produce line lists for the XH
species, i.e. catalogues of transition frequencies $\tilde{\nu}_{ij}$ and
Einstein coefficients $A_{ij}$. The computed line lists for BeH, $^{24}$MgH, $^{25}$MgH, $^{26}$MgH, and $^{40}$CaH
are given in the supplementary materials. The same `spectroscopic' PEC of MgH corrected for BOB was used
to produce  the line lists for all three isotopologues of this species.
The coverage in terms of the frequency range and quantum numbers is illustrated in Table~\ref{tab:ranges}.

The line list file, see Table~\ref{t:Tfile}, contains transition
frequencies $\tilde{\nu}_{ij}$ (\cm), Einstein coefficients $A_{ij}$ ($1/s$),
lower ($i$) state term values $\tilde{E}_{i}^{\prime\prime}$ (\cm), and upper
and lower state quantum numbers for each line. Quantum numbers  are defined
in accordance with the Hund's case (b) formulation: $N$ specifies the total angular
momentum excluding spin, $J=N\pm 1/2$ specifies the total angular momentum
including electron spin, and $e/f$ is the rotationless parity: the
sign of the total parity is defined by $(-1)^{N}$ for a $\Sigma^+$ electronic state so states with $J=N+1/2$ are $e$ states and those with $J=N-1/2$ are $f$ states \citep{75BrHoHu.parity}; and the vibrational quantum number $v$.
Since we ignore hyperfine effects due to the interaction with the nuclear spins, $J$ is assumed to be a conserved
quantity. The $e/f$ labels are omitted from the BeH line list as we
consider the corresponding energies to be  degenerate. The energy levels are
given relative to the corresponding zero point energy $E_{N=0,v=0}^e$.

The spin-rotational coupling plays an important role in the case of
ro-vibrational spectra both of MgH and CaH, in contrast to BeH, where the
fine structure can be considered negligible. The program Level is not capable
of treating the spin-rotation coupling, and therefore the corresponding
ro-vibrational energies of MgH and CaH were computed with the program
DPotFit. Only the Einstein coefficients were generated using Level, i.e.
without spin-rotation effects included. We assume that the $e$--$f$
intensities are negligible, while the transition moments for $e$--$e$ and
$f$--$f$ transitions are equal (see also discussion in the book by
\cite{Herzberg_89}). This can be readily verified for MgH: the known
experimental $e$--$f$ transitions  are indeed very weak \citep{04ShApGo.MgH},
and for CaH no such transitions were reported experimentally, probably for
the same reason. It should be noted that the Level program computes the line
strengths  as `singlet-singlet' transitions and the following correction
factors had to be applied \citep{08Watson.Honl} in order to account for the
fine structure in the MgH and CaH transitions:
\begin{eqnarray}
S^{R}(f\to f) &=& S^{R}_{\rm Level} \frac{2 J^{\prime\prime} + 1}{2 J^{\prime\prime} + 2}, \label{e:R-f-f} \\
S^{P}(f\to f) &=& S^{P}_{\rm Level} \frac{2 J^{\prime\prime} - 1}{2 J^{\prime\prime}}, \label{e:P-f-f} \\
S^{R}(e\to e) &=& S^{R}_{\rm Level} \frac{2 J^{\prime\prime} + 3}{2 J^{\prime\prime} + 2}, \label{e:R-e-e} \\
S^{P}(e\to e) &=& S^{P}_{\rm Level} \frac{2 J^{\prime\prime} + 1}{2 J^{\prime\prime}}. \label{e:P-e-e}
\end{eqnarray}
Here $S^{R/P}$ is the line strength of the $R/P$ transition.

The integrated absorption coefficient $I(\omega_{ij})$ for a $i\to j$
transition can be calculated for a specific temperature $T$ using
\begin{equation}
\label{e:intensity}
I(\omega_{ij}) =\frac{\alpha A_{ij} g_j}{Z(T)\omega_{ij}^2} \left( e^{-E^{\prime\prime}/k_B T} - e^{-E^{\prime}/k_B T}\right),
\end{equation}
where $g_j$  is the degeneracy of the final state,
$Z(T)$ is the partition function, $k_{\rm B}$ is the Boltzmann constant,
$E^{\prime}$ is the energy of the higher state, $E^{\prime\prime}$ is the
energy of the lower state, $\omega$ (\cm) is the
transition wavenumber, and $\alpha$ is the factor $1.3271005 \times10^{-12}\
\mathrm{s}\ \mathrm{cm}^{-2}\ \mathrm{molecule}^{-1}$ so that $I$ has units
of cm/molecule.
For the $X\,{}^2\Sigma$ state of BeH, where the states $i$ are doubly degenerate,
the partition function, $Z(T)$,  is given by:
\begin{equation}\label{e:partfuncS}
    Z(T) = \sum_{i}  g_i  e^{ -E_i/k_{\rm B}T } =
\sum_{i}  g_{\rm ns} (2S+1) (2 N_i+1) e^{ -E_i/k_{\rm B}T },
\end{equation}
where $S=1/2$ is the total electronic spin angular momentum, $g_{\rm ns}=8$ is the nuclear
statistical weight of BeH, and $N_i$ is the rotational angular momentum of the nuclei of the $i$th state. 
Here we follow the HITRAN \citep{HITRAN-A} convention and consider the
nuclear statistical weight of the molecule explicitly in $Z(T)$.
In the case of MgH and CaH in their ground electronic states the
spin-rotational splitting is resolved and $Z(T)$ is given by
\begin{equation}\label{e:partfunc}
    Z(T) = \sum_{i}  g_{\rm ns}  (2 J_i+1) e^{ -E_i/k_{\rm B}T }.
\end{equation}
Here the nuclear statistical weights $g_{\rm ns}=2$ for $^{40}$CaH, $^{24}$MgH and $^{26}$MgH, and
$g_{\rm ns}=12$ for $^{25}$MgH. The nuclear spins of $^{25}$Mg and Be used to define
these values are $5/2$ and $3/2$, respectively, whilst the nuclear
spins of $^{24}$Mg, $^{26}$Mg, and $^{40}$Ca are 0.

\begin{table}
\caption{\label{t:Tfile} Extracts from the line list files for MgH, CaH, and BeH.}
\begin{center}
\scriptsize \tabcolsep=5pt
\renewcommand{\arraystretch}{1.0}
\begin{tabular}{rrrrrrrrr}
\hline
\hline
        1    &  2    &  3        &  4    &         5  & 6 & 7 & 8 & 9      \\
\hline
\multicolumn{9}{c}{$^{24}$MgH} \\
91.048711  &  2.142280E-01  &    320.031578  &   8  &  f  &  0  &   7    &   f   &   0   \\
91.073979  &  2.430318E-01  &    320.227037  &   8  &  e  &  0  &   7    &   e   &   0   \\
91.972295  &  3.231341E-01  &   4472.543625  &   9  &  f  &  3  &   8  &   f   &   3   \\
91.993639  &  3.614000E-01  &   4472.733852  &   9  &  e  &  3  &   8  &   e   &   3   \\
92.716937  &  3.674842E-01  &   6900.960666  &  10  &  f  &  5  &   9  &   f   &   5   \\
92.734931  &  4.063714E-01  &   6901.144783  &  10  &  e  &  5  &   9    &   e   &   5   \\
92.839961  &  1.965889E-01  &   9918.094815  &  14  &  f  &  8  &  13   &   f   &   8   \\
92.846033  &  2.111897E-01  &   9918.247096  &  14  &  e  &  8  &  13   &   e   &   8   \\
92.884148  &  3.860760E-03  &  10261.123179  &   1  &  e  &  11 &   2    &   e   &   10  \\
92.893287  &  2.144867E-03  &  10261.112833  &   1  &  f  &  11 &   2    &   f   &   10  \\
\hline
   \multicolumn{9}{c}{$^{40}$CaH} \\
   15.718268  &  3.013840E-03  &   3673.267366  &   2  &  f  &  3  &  1  &   f   &   3   \\
   15.757038  &  5.424912E-03  &   3673.325586  &   2  &  e  &  3  &  1  &   e   &   3   \\
   16.110939  &  2.957107E-03  &   2490.047709  &   2  &  f  &  2  &  1  &   f   &   2   \\
   16.132829  &  6.775908E-03  &  13147.887247  &   3  &  f  &  13   &  2  &   f   &   13  \\
   16.151200  &  5.322792E-03  &   2490.108168  &   2  &  e  &  2  &  1  &   e   &   2   \\
   16.154575  &  6.453246E-03  &  13147.941920  &   3  &  e  &  13   &  2  &   e   &   13  \\
   16.500092  &  2.878773E-03  &   1268.370045  &   2  &  f  &  1  &  1  &   f   &   1   \\
   16.541842  &  5.181792E-03  &   1268.432737  &   2  &  e  &  1  &  1  &   e   &   1   \\
   16.887199  &  2.782460E-03  &  8.434990  &   2  &  f  &  0  &  1  &   f   &   0   \\
   16.930436  &  5.008428E-03  &  8.499915  &   2  &  e  &  0  &  1  &   e   &   0   \\
\hline
\multicolumn{9}{c}{$^9$BeH} \\
   19.079220  &  1.306725E-04  &   3896.871469  &   1  &   &  2  &  0    &   &   2   \\
   19.707403  &  9.607023E-05  &   1986.417648  &   1  &   &  1  &  0    &   &   1   \\
   20.327181  &  6.337485E-05  &      0.000000  &   1  &   &  0  &  0    &   &   0   \\
   22.576116  &  6.097568E-04  &  15721.783485  &   2  &   &  10   &  1    &   &   10  \\
   23.667196  &  2.713953E-04  &  16426.182379  &   3  &   &  11   &  2    &   &   11  \\
   26.231114  &  1.246778E-03  &  14731.961064  &   2  &   &  9  &  1    &   &   9   \\
   28.814926  &  1.753481E-03  &  13541.079476  &   2  &   &  8  &  1    &   &   8   \\
   30.852363  &  2.056734E-03  &  12198.490607  &   2  &   &  7  &  1    &   &   7   \\
   30.856982  &  5.807508E-04  &  16449.849576  &   4  &   &  11   &  3  &   &   11  \\
   32.574171  &  2.147658E-03  &  10732.929225  &   2  &   &  6  &  1    &   &   6   \\
\hline
\end{tabular}

\begin{tabular}{cll}
\hline
             Column           &    Notation                 &      \\
\hline
1 &  $\tilde{\nu}_{ij}$         & Transition frequency in \cm \\
2 &  $A_{ij}$                   & Einstein A coefficient in s$^{-1}$\\
3 &  $\tilde{E}^{\prime\prime}$ & lower state energy in \cm \\
4 &  $N^{\prime}$               & upper state rotational quantum number\\
5 &  $p^{\prime}$               & upper state rotationless parity\\
6 &  $v^{\prime}$               & upper state vibrational quantum number\\
7 &  $N^{\prime\prime}$         & lower state rotational quantum number\\
8 &  $p^{\prime\prime}$         & lower state rotationless parity\\
9 &  $v^{\prime\prime}$         & lower state vibrational quantum number\\
\hline
\hline
\end{tabular}

\end{center}

\end{table}


\begin{table}
\begin{center}

 \caption{\label{tab:ranges} The completeness and coverage of our  line lists for BeH, MgH, and CaH. }
\footnotesize
\begin{tabular}{lrrrrr}
\hline
                        & $^9$BeH                & $^{24}$MgH& $^{25}$MgH& $^{26}$MgH& $^{40}$CaH        \\
\hline
$\nu_{\rm max}$ (\cm)   &            16\,400 &    10~354 &    10~354 &    10~355 &    15\,278 \\
$v_{\rm max}$           &                 11 &        11 &        11 &        11 &       19   \\
$N_{\rm max}$           &                 59 &        60 &        60 &        61 &       74   \\
$N_{\rm lines}$         &               3968 &      6716 &      6751 &      6754 &    26\,980 \\
\hline
\end{tabular}
\end{center}
\end{table}

\begin{figure}
\caption{
\label{fig:comps}
Theoretical stick absorption spectra of BeH, $^{24}$MgH, and $^{40}$CaH at the room temperature compared
to the experimental data reported by
\protect\cite{03ShTeBe.BeH},
\protect\cite{04ShApGo.MgH}, and
\protect\cite{04ShWaGo.CaH}, respectively.
The experimental spectra are scaled to match the
theoretical intensities.
}
\begin{center}
{\includegraphics{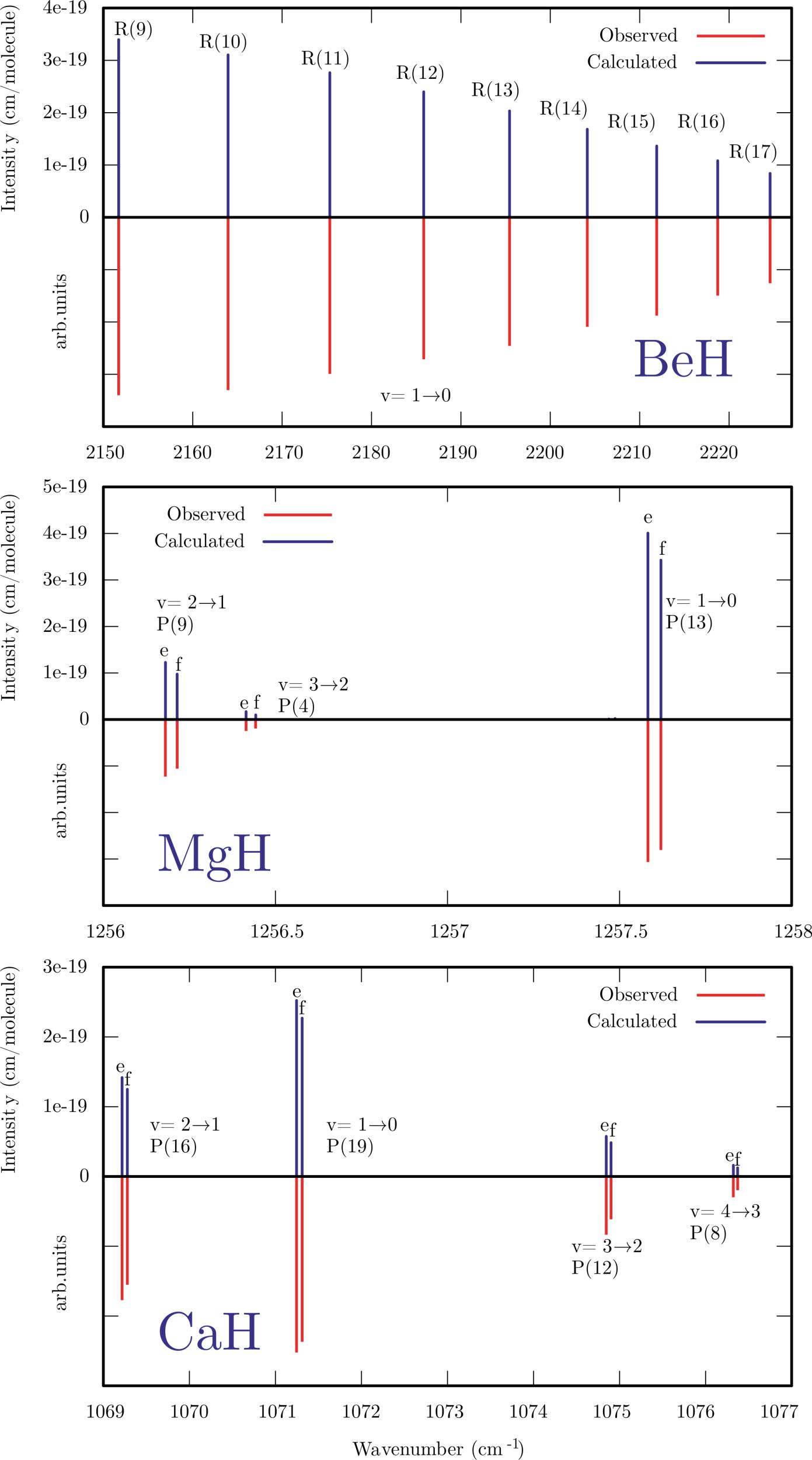}} \\
\end{center}
\end{figure}

\begin{figure}
\begin{center}
{\includegraphics{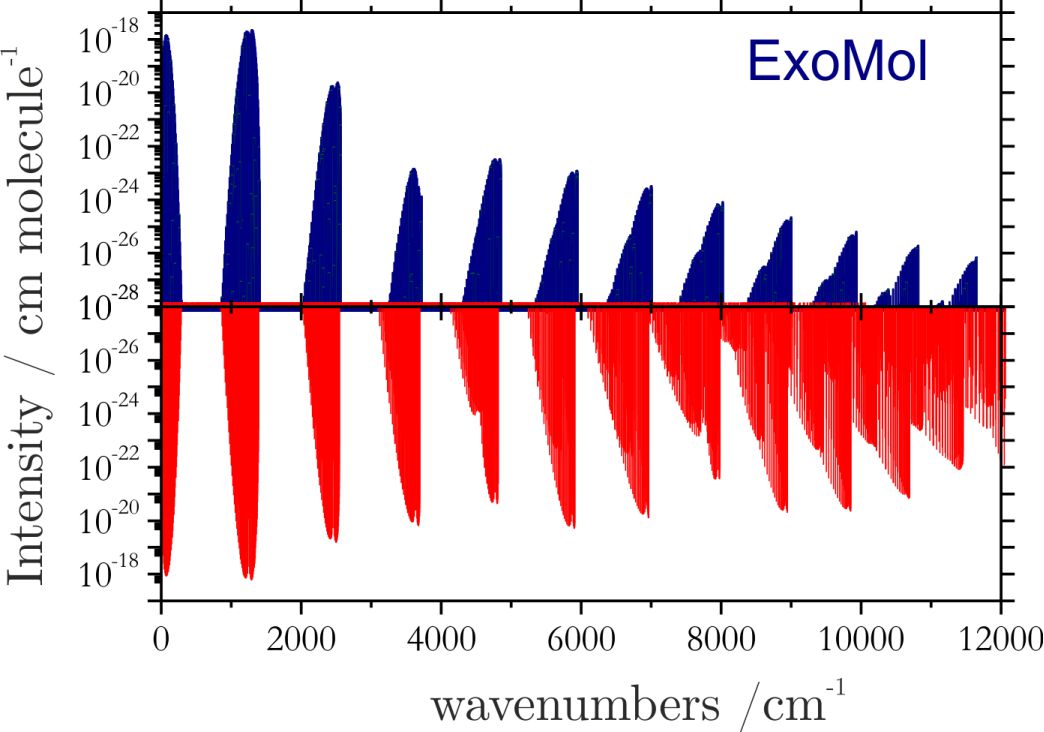}}
\caption{Absorption spectra of CaH at T=296~K: ExoMol vs UGAMOP}
\label{fig:cah:Exomol-vs-UGAMOP:300K}
\end{center}
\end{figure}

\begin{figure}
\begin{center}
{\includegraphics{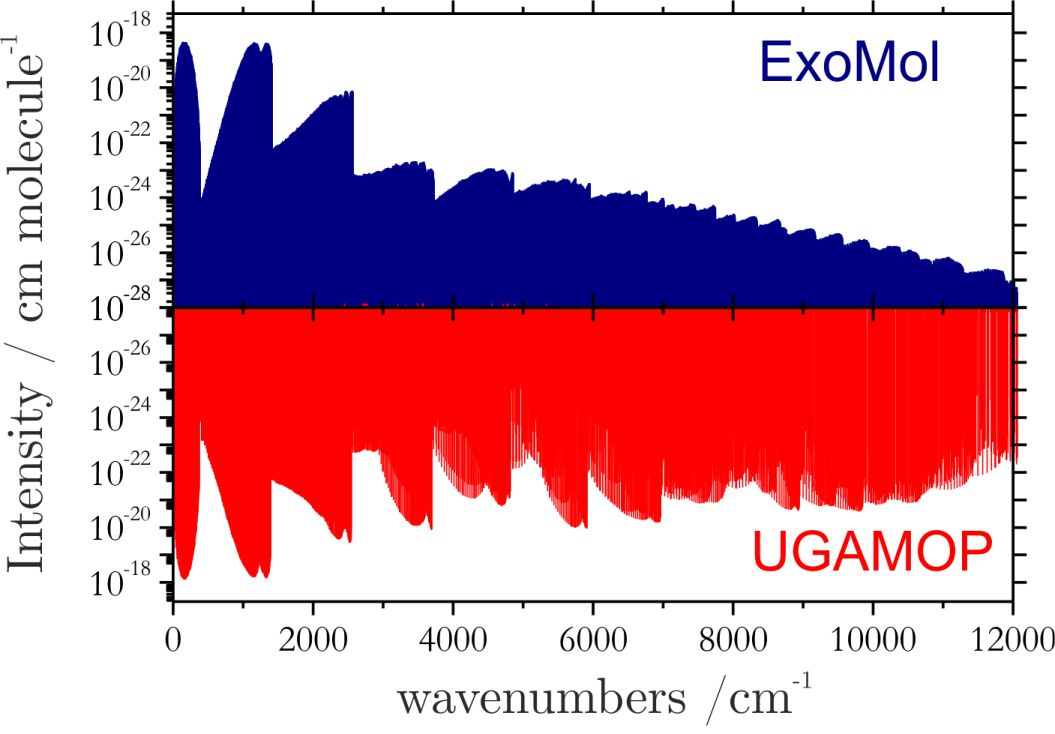}}
\caption{Absorption spectra of CaH at T=1500~K: ExoMol vs UGAMOP.}
\label{fig:cah:Exomol-vs-UGAMOP:1500K}
\end{center}
\end{figure}

\begin{figure}
\begin{center}
{\includegraphics{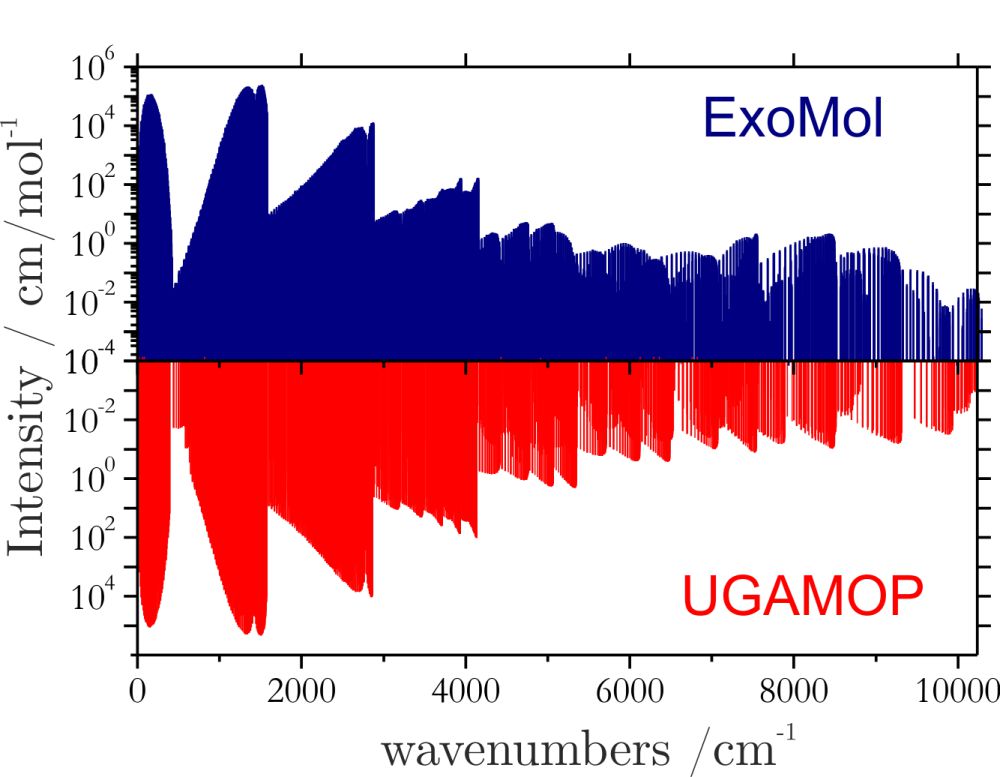}}
\caption{Absorption spectra of MgH at T=1500~K: ExoMol  vs UGAMOP.}
\label{fig:mgh:Exomol-vs-UGAMOP:1500K}
\end{center}
\end{figure}

\begin{figure}
\begin{center}
{\includegraphics{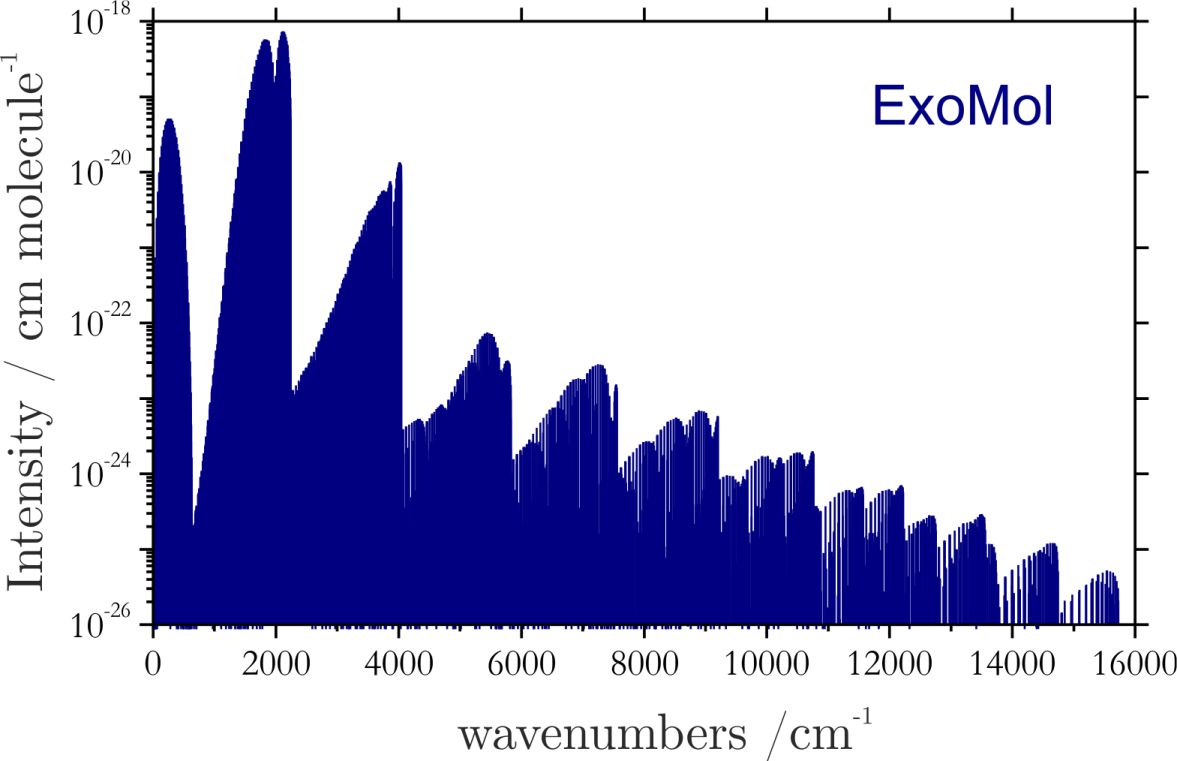}}
\caption{ExoMol absorption spectra of BeH at T=1500~K.}
\label{fig:beh:300K}
\end{center}
\end{figure}

High resolution comparisons with specific regions of the experimental spectra
for each molecule are given in Fig. \ref{fig:comps}; note that the
experimental spectrum has been scaled to the theoretical data by matching one
line since none of the reported measurements are absolute. These comparisons
demonstrate the accuracy of our procedure. In
Figs.~\ref{fig:cah:Exomol-vs-UGAMOP:300K},
\ref{fig:cah:Exomol-vs-UGAMOP:1500K} and \ref{fig:mgh:Exomol-vs-UGAMOP:1500K}
we also compare synthetic absorption spectra of CaH and MgH at the room
temperature as well as at $T=1500$~K generated using our and the UGAMOP line
lists. The line intensities in these figures are given by sticks. For a
better comparison the UGAMOP line intensities are divided by 2 in order to
account for the difference in treating the fine (i.e. spin-rotational)
splitting. In the case of CaH our intensities in the higher frequency range
are systematically weaker than that of UGAMOP. Fig.~\ref{fig:beh:300K}
illustrates a 1500~K absorption spectrum for BeH.

\subsection{Partition function}

We used the calculated energies of all three molecules to generate the
partition function values for a range of temperatures from 0 to 3000~K. These
values were represented by the following polynomial expansion (\cite{jt263})
via a least-squares fit:
\begin{equation}\label{e:Q:a}
    \log(Z)= \sum_{n=1}^{6} a_i \log(T)^i.
\end{equation}
The resulting expansion constants $a_i$ for BeH, $^{24}$MgH, and $^{40}$CaH are listed in
Table~\ref{t:Q:a}. To a very good approximation the partition functions
of $^{24}$MgH, $^{25}$MgH, and $^{26}$MgH are in the ratio 2:12:2, which takes
account of the increased nuclear spin statistical weight of $^{25}$MgH. 
This scaling can be used to obtain
$Z(T)$ for the latter two species.

We have also compared our $Z(T)$ of CaH to the partition
function values computed using the UGAMOP energies of CaH. Even though the
difference in our and UGAMOP energies is large
(Table~\ref{tab:CaH:energies}), the partition functions agree within 3\% for
the whole range of $T$ up to 3000~K.

\cite{81Irxxxx.partfunc} and \cite{84SaTaxx.partfunc} reported similar
expansions of $Z(T)$ for a large set of molecules, which we can use to
validate our results. For example, for $^{24}$MgH at $T=2000$K from
\cite{81Irxxxx.partfunc} and \cite{84SaTaxx.partfunc} one obtains $Z=659.8$
and $Z=783.2$, respectively to compare with our value $Q/g_{\rm s}  = 811.0$.
The factor $g_{\rm s}$ = 2 here is needed to account for the nuclear spin statistics
absent in the approach used  by \cite{81Irxxxx.partfunc} and
\cite{84SaTaxx.partfunc} (see discussion in \cite{ExoMol0}). However at room temperature
even a small difference in $Z(T)$ becomes important. Our value at $T=296$~K is
72.8, while \cite{84SaTaxx.partfunc} give 65.7. These values are also
collected in Table~\ref{t:Q:a} together with corresponding $Z(T)$ obtained
for CaH and BeH. Note a striking difference of about 50\% for $Z$(296K) of
CaH between our value and that from \cite{84SaTaxx.partfunc}.

\begin{table}
\caption{\label{t:Q:a}  BeH, $^{24}$MgH, and $^{40}$CaH partition function,
$Z(T)$, expansion parameters for Eq.~\eqref{e:Q:a} and sample partition function values.
These values are compared to $Z(T)$ obtained using parameters from UGAMOP and \protect\cite{84SaTaxx.partfunc}.
Please note that our and UGAMOP's values are computed using the nuclear statistical factors 8, 2, and 2 for BeH, $^{24}$MgH, and $^{40}$CaH,
respectively.  }
\begin{center}
\begin{tabular}{lrrrr}
\hline
   Parameter & &             BeH  &    $^{24}$MgH  &    $^{40}$CaH    \\
\hline
 $ a_0  $   &&           -3.69(26)  &     -4.73(42)  &     -2.38(36)    \\
 $ a_1  $   &&           14.63(78)  &      16.1(12)  &       9.2(11)    \\
 $ a_2  $   &&          -17.95(91)  &     -19.8(14)  &     -11.5(13)    \\
 $ a_3  $   &&           11.64(54)  &     13.01(86)  &      8.01(75)    \\
 $ a_4  $   &&           -4.01(18)  &     -4.61(28)  &     -2.98(25)    \\
 $ a_5  $   &&           0.703(30)  &     0.836(48)  &     0.564(42)    \\
 $ a_6  $   &&         -0.0490(21)  &   -0.0606(33)  &   -0.0423(29)    \\
\hline
 & &      BeH  &    $^{24}$MgH  &    $^{40}$CaH  \\
\hline
$g_{\rm ns}$                        &                  &         8  &         2   &          2   \\
$Z$(296K) $\times g_{\rm ns}$       &   S\&T$^a$       &      93.9  &     129.5   &       93.9   \\
$Z$(296K)                           &   UGAMOP         &            &     147.8   &      195.9   \\
$Z$(296K)                           &   This work      &     335.0  &     145.6   &      199.9   \\
$Z$(2000K) $\times g_{\rm ns}$      &   S\&T$^a$       &    2276.9  &    1566.4   &     2276.9   \\
$Z$(2000K)                          &   UGAMOP         &            &    6575.5   &     2451.7   \\
$Z$(2000K)                          &   This work      &    3019.5  &    1622.4   &     2253.5   \\
\hline
\end{tabular}
\end{center}
$^a$ \cite{84SaTaxx.partfunc}

\end{table}

\section{Conclusions}

Our potentials are highly accurate for low vibrational levels
but there is undoubtedly some loss of accuracy for higher vibrational
states given the lack of experimental data to constrain the curves.
For BeH,  high accuracy should  be
achieved for all transitions involving states up to $v = 11$.
For MgH, we can be confident of sub-wavenumber
accuracy for $v>4$ levels as these higher levels can be characterised using
electronic transitions. For CaH, it is not possible to test any of
the vibrational states higher than $v=4$ since there is no
experimental data.

Line lists for the rotation-vibration transitions within the
ground states of the BeH, $^{24}$MgH, $^{25}$MgH and $^{26}$MgH, and $^{40}$CaH
molecules are presented which should be
accurate over an extended temperature range. These line lists were computed
using potential energy curves fitted to experimental data and {\it ab initio}
dipole moment curves obtained at high levels of theory. A complete line list
for each of these molecules can be downloaded from the CDS, via
\url{ftp://cdsarc.u-strasbg.fr/pub/cats/J/MNRAS/}, or via
\url{http://cdsarc.u-strasbg.fr/viz-bin/qcat?J/MNRAS/}. The line lists
together with auxiliary data including the potential parameters, dipole
moment functions, and theoretical energy levels can be also obtained at
\url{www.exomol.com}.

\section*{Acknowledgements}

This work is supported by ERC Advanced Investigator Project 267219. We also
thank the present members of the ExoMol team, Ala'a Azzam, Bob Barber,
Lorenzo Lodi, Andrei Patrascu, Oleg Polyansky, and Clara
Silva Sousa, as well as Attila Cs\'{a}sz\'{a}r for many helpful discussions
on the project and Olga Yurchenko for her help.

\bibliographystyle{mn2e}

\label{lastpage}

\end{document}